\documentclass[twocolumn,a4paper,showpacs,pra,aps]{revtex4-1}

\usepackage[dvipsnames]{xcolor}
 \usepackage{amsmath}
 \usepackage{amssymb}
 \usepackage{bbold}
 \usepackage{latexsym}
 \usepackage{amsfonts}
 \usepackage[caption=false]{subfig}
 \usepackage{epsfig}
 \usepackage{psfrag}
 \usepackage{color}
 \usepackage{rotating}
 \definecolor{darkblue}{rgb}{0,0,.5}
 \usepackage[linktocpage, colorlinks=true ,linkcolor=darkblue, citecolor=darkblue]{hyperref}
 \usepackage[all]{hypcap}
 
\usepackage{graphicx}
\usepackage{dcolumn}
\usepackage{bbm}
\usepackage{bm}
\usepackage{overpic}
  \newcommand{\ket}[1]{\left|#1\right>}
  \newcommand{\bra}[1]{\left<#1\right|}
  \newcommand{\expval}[1]{\left< #1 \right>}
  \newcommand{\braket}[2]{\left<#1|#2\right>}

\begin{document}

%%%%%%%%%%%%%%%%%%%%%%%%%%%%%%%%%%%%%%%%%%%%%%%%%%%%%%%%%%%%%%%%
\title{Critical quasienergy states in driven many-body systems}
%%%%%%%%%%%%%%%%%%%%%%%%%%%%%%%%%%%%%%%%%%%%%%%%%%%%%%%%%%%%%%%%

\author{V. M. Bastidas$^1$}
\email{victor@physik.tu-berlin.de}

\author{G. Engelhardt$^{1}$} 

\author{P. P{\'e}rez-Fern{\'a}ndez$^{1,2}$}

\author{M. Vogl$^{1,3}$}

\author{T. Brandes$^1$}
\affiliation{$^1$Institut f\"ur Theoretische Physik, Technische Universit\"at 
Berlin, Hardenbergstr. 36, 10623 Berlin, Germany}
\affiliation{$^2$Departamento de F\'isica Aplicada III, Escuela Superior de 
Ingenier{\'i}a, 
Universidad de Sevilla, Camino de los Descubrimientos s/n, ES-41092 Sevilla,
Spain}
\affiliation{$^3$ Max-Planck-Institut f\"ur Physik komplexer Systeme, 
N\"othnitzer Stra\ss e 38, 01187 Dresden, Germany}

%
%%%%%%%%%%%%%%%%%%%%%
\begin{abstract}
%%%%%%%%%%%%%%%%%%%%%
%
We discuss singularities in the spectrum of driven many-body spin systems.
In contrast to undriven models, the driving allows us to control the geometry of 
the quasienergy landscape. 
As a consequence, one can engineer singularities in the density of
quasienergy states by tuning an external control.
We show that the density of levels exhibits logarithmic divergences at the saddle points, while
jumps are due to local minima of the quasienergy landscape.
We discuss the characteristic signatures of these divergences in observables like the magnetization, which should be measurable with current technology.
%
%%%%%%%%%%%%%%%%%%%
\end{abstract}
%%%%%%%%%%%%%%%%%%%
%


%%%%%%%%%%%%%%%%%%%%%%%%%%%%%%%%%%%%%%%%%%%%%%%%%%%%%%%%%%%%%%%%%%%%%%%%%%%%%%%%
\maketitle
%%%%%%%%%%%%%%%%%%%%%%%%%%%%%%%%%%%%%%%%%%%%%%%%%%%%%%%%%%%%%%%%%%%%%%%%%%%%%%%%

%
%%%%%%%%%%%%%%%%%%%%%%%
\section{Introduction}
\label{sec:I}
%%%%%%%%%%%%%%%%%%%%%%
%
A quantum phase transition (QPT) 
is characterized by non-analytical behavior of the ground-state properties of 
the system, when a control parameter 
crosses the quantum critical point~\cite{Sachdev}. 
Rather recently 
it has been shown that quantum criticality can appear also in excited states of the system, which is referred to as an excited-state quantum phase transition 
(ESQPT)~\cite{Cejnar06, Caprio08, CejnarStransky08, Leyvraz05}.
This kind of quantum criticality can be found in a wide variety of models 
in different communities, which range from nuclear physics, 
with the interacting Boson~\cite{Arias03, CejnarRMF} and the 
Lipkin-Meshkov-Glick  (LMG) models~\cite{RibeiroESQPT}, to quantum monodromy in molecular physics~\cite{monodromy}  and the Dicke and 
Jaynes-Cummings models in quantum optics~\cite{Brandes,BrandesESQPT,PPF11E, PPF11A}. 

ESQPTs can induce dramatic effects on the quantum dynamics of the system. 
For example, environments with ESQPTs enhance decoherence on quantum registers, which has implications for quantum computation~\cite{PPF08A}. In addition, thermalization processes can be affected by ESQPTs due to degeneracies in the spectrum~\cite{Puebla}.

Most of the aforementioned models exhibit a ESQPT, that leads to a logarithmic singularity in the density of states~\cite{Caprio08}.
Such a singularity occurs at a critical energy, and it is a quantum manifestation of
the separatrix, i.e., a homoclinic or heteroclinic orbit of the corresponding semiclassical model.
%
% 
%%%%
\begin{figure}
\includegraphics[clip=true,width=0.9\linewidth]{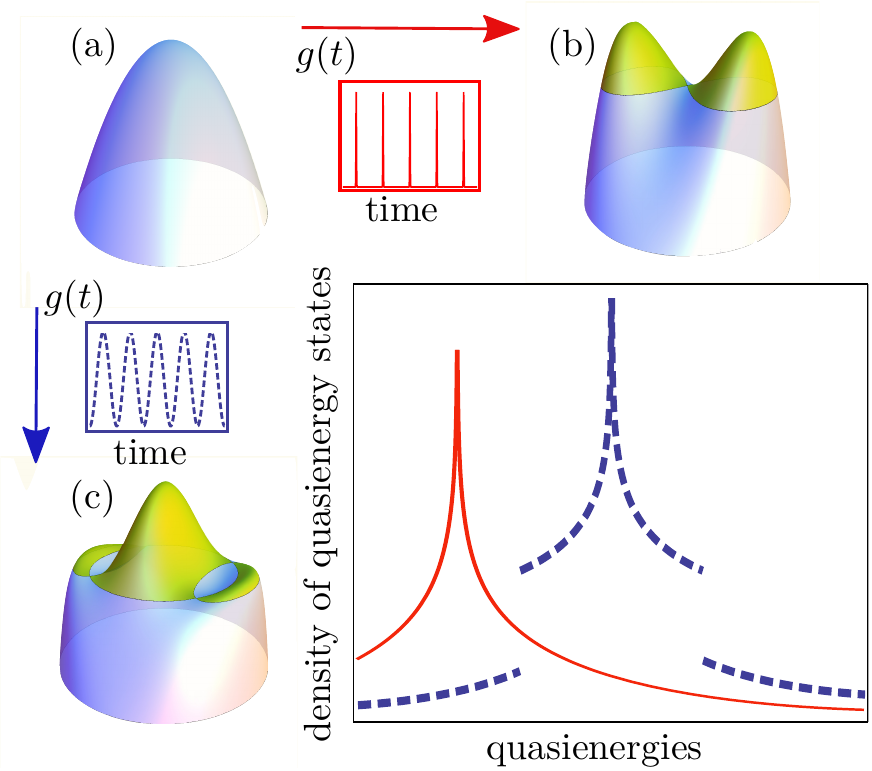}
\caption{\label{Fig0} (Color online)
    Applying an external control $g(t)$ to the undriven system (a) allows to engineer geometrical features of the quasi-energy landscape, which leaves measurable signatures in observables.
    Depending on the shape of the external control, e.g., delta-kick-type (solid red) or monochromatic (dashed blue), the emergent saddle points are connected via homo- (b) or hetero-clinic orbits (c), giving rise
    to characteristic density of quasienergy states.}
\end{figure}
%%%%%
%

Observing such a separatrix experimentally is an active field for non-driven models.
Recently, the classical bifurcation in the anisotropic LMG model has been observed in Bose-Einstein condensates~\cite{Zibold2010,Gross2010}. 
Furthermore, in the context of spinor Bose-Einstein condensates~\cite{Stamper-KurnUeda,KawaguchiUeda}, quantum signatures of a semiclassical
separatrix have been realized experimentally~\cite{Gerving2012,Hamley2012}. 
In these experiments, a Gaussian initial state is prepared at the hyperbolic fixed point of the separatrix, where the subsequent evolution leads to non-Gaussian states, and the creation of spin squeezing~\cite{KitagawaUeda,NoriSpinSqueez}.

On the other hand, periodically driven systems -- as depicted in Fig.~\ref{Fig0} -- have been proven to be a seminal
playground in both theoretical and experimental manner. 
Starting from the possibility to create effective interactions, it is possible to control 
the topological states of matter~\cite{Oka2009,Lindner2011,Inoue2010,Jiang2011,Reynoso2013,Kitagawa,Grushin},
create unconventional phases in the 
Dicke, LMG and  the Ising models~\cite{Bastidas2012,Hangui, Engelhardt2013, Bastidas4},
and suppress coherent tunneling in
a two-mode Bose-Hubbard model~\cite{Hangui}. 

Since there is currently a rising interest in the experimental investigation of driven
mean-field-type spin models~\cite{Hoang,Zhao,Chaudhury1,Chaudhury2}, it 
 is natural to ask if signatures analogous to ESQPTs in undriven systems can also
be found in driven systems, where the energy is not conserved and it is not possible to define either a ground state or excited states. 

In this paper we give a step in this direction, and
develop a general formalism to calculate analytically 
the density of quasienergy states (DOQS) under the stationary-phase approximation~\cite{HaakeBuch}.  To illustrate this, we show in Fig.~\ref{Fig0} how driving can generate a separatrix leading to characteristic features in the DOQS. 
Specifically, we apply this method to the well-known kicked top model~\cite{HaakeBuch,KickedTopESQPT} and to the ac-driven LMG model~\cite{Hangui,Engelhardt2013}. 

Concerning  the kicked top,  in a previous work, we found signatures of quantum criticality in its spectrum in the regular regime~\cite{KickedTopESQPT}.  A recent paper describe a method to improve the convergence of the effective Hamiltonian~\cite{Sarkar}.
However, it is an open question whether quantum criticality and a convergent effective Hamiltonian could 
be still found in the chaotic regime~\cite{HaakeBuch}.

The rest of the paper is organized as follows. 
In Sec.~\ref{sec:II} we introduce the general theory, including a very short introduction to Floquet theory~(\ref{sec:IIa}), the definition of the DOQS~(\ref{sec:IIb}), 
the bosonization procedure~(\ref{sec:IIc}) and the actual calculation of the DOQS~(\ref{sec:IId}).
In Sec.~\ref{sec:III} we then apply the theory to two different models. We introduce their effective Hamiltonians~(\ref{sec:IIIa}), discuss  the quasienergy landscapes~(\ref{sec:IIIb}) and the corresponding 
critical quasienergy states~(\ref{sec:IIIc}). Last, we look at experimentally accessible signatures of these states~(\ref{sec:IIId}).
The conclusion~(\ref{sec:IV}) is followed by the appendix, containing some more detailed calculations.

\section{General theory}
\label{sec:II}
In this section we introduce the general idea. We first discuss the basics of Floquet theory and how to obtain effective Hamiltonians, then we look at 
a suitable formulation for the DOQS in this context. Next, we discuss the bosonization of mean-field type models and show how this allows one
to obtain analytic results for the DOQS.
%%%%%%%%%
\subsection{Floquet theory and effective Hamiltonian}
\label{sec:IIa}
%%%%%%%%%
%
In this paper we discuss quantum criticality arising in excited states 
of time periodic Hamiltonians $H(t)=H(t+T)$ describing mean-field-type manybody systems.
Throughout this paper, $T=2\pi/\Omega$ denotes the period and $\Omega$ the frequency of the external driving. 
Due to the periodicity of the Hamiltonian, it is convenient to use Floquet theory~\cite{Shirley,Sambe} to describe the quantum 
evolution of the system.
For this purpose, we use the Floquet operator, which is the evolution operator in one period of the external driving
%%%
\begin{equation}
      \label{FloquetOperator}
            \hat{\mathcal{F}}=\hat{U}(T) =\hat{\mathcal{T}}\exp\left[-\mathrm{i}\int_{0}^{T}\hat{H}(\tau) \ d\tau\right]
      ,
\end{equation}
%%%
where $\hat{\mathcal{T}}$ is the time-ordering operator. 
The Floquet modes $\ket{\Phi_{\mu}(t)}= \ket{\Phi_{\mu}(t+T)}$ are obtained by solving the eigenvalue problem  
%%%
\begin{equation}
      \label{FloquetEVProblem}
            \hat{\mathcal{F}}\ket{\Phi_{\mu}(0)} =e^{-\mathrm{i} \varepsilon_{\mu} T} \ket{\Phi_{\mu}(0)}
      ,
\end{equation}
%%%
where $\varepsilon_{\mu}$ are the quasienergies~\cite{HanggiGrifoni}.

In contrast to undriven systems, the energy is not conserved under external driving. 
Correspondingly,  quasienergies do not have an intrinsic ordering as  energies do. 
This situation arises because if $\ket{\Phi_{\mu}(0)}$ satisfies Eq.~\eqref{FloquetEVProblem},
there is an infinite set of states $\ket{\Phi_{\mu,n}(t)}=e^{\mathrm{i} n\Omega t}\ket{\Phi_{\mu}(t)}$, such that
\begin{equation}
      \label{AltFloquetEVProblem}
            \hat{\mathcal{F}}\ket{\Phi_{\mu,n}(0)} =e^{-\mathrm{i} \varepsilon_{\mu,n} T} \ket{\Phi_{\mu,n}(0)}
      \ ,
\end{equation}
with quasienergies $\varepsilon_{\mu,n}=\varepsilon_{\mu}+n\Omega$. 
Due to the lack of ordering of the quasienergies, throughout the paper we consider only the first Brillouin zone, which is defined by $- \Omega/2\leq \varepsilon_{\mu} \leq \Omega/2$.

The Floquet operator allows one to describe the system 
stroboscopically~\cite{Shirley,Sambe,HanggiGrifoni, HaakeBuch}. 
That is, given an initial state $\ket{\Psi(0)}=\sum_{\mu}c_{\mu}\ket{\Phi_{\mu}(0)}$, the state at discrete times $t_{m}=mT$ is given by
\begin{equation}
      \label{EvolutionStroboscopic}
            \ket{\Psi(mT)}=\hat{\mathcal{F}}^{m}\ket{\Psi(0)} 
            =\sum_{\mu}c_{\mu}e^{-\mathrm{i} m\varepsilon_{\mu} 
T}\ket{\Phi_{\mu}(0)}
 ,
\end{equation}
which resembles the evolution operator for a time-independent Hamiltonian~\cite{Kitagawa,Hangui,Engelhardt2013,Bastidas2012, 
Bastidas4}. 
This motivates the introduction of an effective Hamiltonian (EH)  
$\hat{H}_{E}$ for the system, such that 
$\hat{\mathcal{F}}=e^{-\mathrm{i}\hat{H}_{\text{E}}T}$. 
Thus, this EH will generate stroboscopic dynamics.

Following the definition of the EH, it is clear  that the Floquet modes 
satisfy $\hat{H}_E \ket{\Phi_{\mu}(0)}=E_\mu \ket{\Phi_{\mu}(0)}$,
where $\{E_\mu\}$ are the 
\textit{unfolded} quasienergies, as it is discussed in Ref.~\cite{KickedTopESQPT}. 
In contrast to the genuine ones $\{\varepsilon_{\mu}\}$, they obey an intrinsic ordering.
Furthermore, it is possible to map $E_\mu$
onto genuine quasienergies $\varepsilon_{\mu}$ by $\varepsilon_\mu = E_\mu 
\ \text{mod} \  \Omega$.

It is worth noticing that unfolded quasienergies are analogous to the energies 
of an undriven system. 
Therefore, by using them, one can take advantage of the knowledge we 
have about QPTs and ESQPTs in undriven systems~\cite{Sachdev,Cejnar06, Caprio08, CejnarStransky08, Leyvraz05} in order to analyze quantum
criticality in driven systems. 
%

%
%%%%%%%%%                                                                  
\subsection{The density of quasienergy states}
\label{sec:IIb}
%%%%%%%%%
%
In this section we develop a general formalism to calculate the DOQS for mean-field-type driven systems. To this end, we assume that we are working in a parameter regime where the EH is well defined, justifying $\hat{\mathcal{F}}=e^{-\mathrm{i}\hat{H}_{\text{E}}T}$.

Unlike for undriven systems, the lack of ordering of the quasienergies $\{\varepsilon_{\mu}\}$ leads to subtleties in the definition of the DOQS.
Similarly to Ref.~\cite{HaakeBuch}, we consider here an alternative representation of the DOQS
%%%
\begin{equation}
      \label{DOQSDef}
            \rho(\varepsilon)=\frac{1}{2\pi}+\frac{1}{\pi M}{\rm Re}\left[\sum_{n=1}^{\infty}\mathcal{T}_{n}e^{\mathrm{i} n \varepsilon T}\right]
       ,
\end{equation}
%%%
where $\mathcal{T}_{n}=\sum_{\mu}e^{-\mathrm{i}n\varepsilon_{\mu}T}=\text{tr}\hat{\mathcal{F}}^{n}$ is the trace of the $n$-th power of the Floquet operator $\mathcal{F}$ defined in Eq.~\eqref{FloquetOperator}, and we have assumed a Hilbert space of dimension $M$ as in the supplementary material of Ref.~\cite{KickedTopESQPT}. 

To calculate the trace, in the following we discuss the bosonization of mean-field-type models~\cite{Klein-Marshalek}, which leads to the definition of the quasienergy landscape (QEL) and enables us to calculate the DOQS analytically.

%
%%%%%%%%%                                                                  
\subsection{Bosonization and the quasienergy landscape}
\label{sec:IIc}
%%%%%%%%%
We begin by assuming that our EH can be written as a function of the generators $\{L_{a}\}$ of a Lie algebra $\mathfrak{g}$~\cite{Klein-Marshalek}. 
Furthermore, we require a representation of $\mathfrak{g}$ in terms of a set of $f$ bosonic operators $\boldsymbol{a}=(a_1,a_2,\ldots,a_f)$. For example, in the case of $\mathfrak{g}=su(2)$, we have $f=1$ if we invoke the Holstein-Primakoff representation~\cite{HolsteinPrimakoffOriginal}
%%%
\begin{align}
	    J_x&=j-a_{1}^{\dagger}a_{1} ,\nonumber \\ 
	    J_{z}+\mathrm{i}J_{y}&=a_{1}^{\dagger}~\sqrt{2j-a_{1}^{\dagger}a_{1}} ,\nonumber \\ 
            J_{z}-\mathrm{i}J_{y}&=\sqrt{2j-a_{1}^{\dagger}a_{1}}~~a_{1} 
             \label{HolsteinPrimakoff}
      \ .
\end{align}
%%%
Provided a convenient bosonic representation of the Lie algebra~\cite{Klein-Marshalek},
the bosonization procedure can be generalized to other mean-field-type systems with higher spin, such as spinor Bose-Einstein condensates~\cite{Stamper-KurnUeda,KawaguchiUeda} or atomic systems coupled to optical cavities, such as the Dicke model~\cite{BrandesESQPT,Brandes} or cavity QED with atoms in $\Lambda$-configuration~\cite{Hayn2011,Hayn2012}.

After bosonization of the EH, we introduce the mean fields $\boldsymbol{\alpha}=(\alpha_1,\alpha_2,\ldots,\alpha_f)$. Formally this can be achieved by using a displacement operator~\cite{Glauber}
\begin{equation}
      \label{DisplacementOperator}
            \hat{D}(\sqrt{\mathcal{N}}\boldsymbol{\alpha})=\exp\left[\sqrt{\mathcal{N}}\left(\boldsymbol{\alpha} \cdotp
\boldsymbol{a}^{\dagger}-\boldsymbol{\alpha}^{*}\cdotp \boldsymbol{a}\right)\right] ,
\end{equation}
%%%
 where $\alpha$ is a complex variational parameter  such that 
%%%
\begin{equation}
      \label{ActionDisplacement}
            \hat{D}^{\dagger}(\sqrt{\mathcal{N}}\boldsymbol{\alpha}) a_{l}  \hat{D}(\sqrt{\mathcal{N}}\boldsymbol{\alpha})=a_{l}+\sqrt{\mathcal{N}}
\alpha_{l}
\end{equation}
%%%
for $l\in\{1,2,\ldots,f\}$. The scaling factor $\sqrt{\mathcal{N}}$ depends on the model and the dimension $M$ of the Hilbert space. In the case  of Hamiltonian~\eqref{SpinDriven}, the dimension of the Hilbert space is $M=2j+1$ and the scaling factor reads $\sqrt{\mathcal{N}}=\sqrt{j}$.

We define the shifted Hamiltonian as
\begin{equation}
      \label{ShiftedHamiltonian}
            \hat{H}^{(\boldsymbol{\alpha})}_{\text{E}}=\hat{D}^{\dagger}(\sqrt{\mathcal{N}}\boldsymbol{\alpha})\hat{H}_{
\text{E}} \hat{D}(\sqrt{\mathcal{N}}\boldsymbol{\alpha})
\end{equation}
and expand it neglecting terms of the order $\mathcal{O}(\mathcal{N}^{-1/2})$
\begin{equation}
      \label{MeanFieldEffHam}
            \hat{H}^{(\boldsymbol{\alpha})}_{\text{E}} 
\approx \mathcal{N} E_{G}(\boldsymbol{\alpha},\boldsymbol{\alpha}^{*})  +\sqrt{\mathcal{N}}\ 
\hat{H}_{\text{E}}^{L}(\boldsymbol{\alpha},\boldsymbol{\alpha}^{*})+\hat{H}_{\text{E}}^{Q}(\boldsymbol{\alpha},\boldsymbol{\alpha}^{*}) 
     ,
\end{equation}
%%% 
where $E_{G}(\boldsymbol{\alpha},\boldsymbol{\alpha}^{\ast})$ denotes the quasienergy landscape (QEL).
The QEL determines features of the quadratic $(\hat{H}_{\text{E}}^{Q})$ and linear $(\hat{H}_{\text{E}}^{L})$ terms in the bosonic operators $\boldsymbol{a}$ and $\boldsymbol{a}^{\dagger}$. For example, the term $\hat{H}_{\text{E}}^{L}$ vanishes
at the critical points where $\frac{\partial}{\partial \alpha_{l}}E_{G}(\boldsymbol{\alpha},\boldsymbol{\alpha}^{*})=\frac{\partial}{\partial \alpha_l^{*}}E_{G}(\boldsymbol{\alpha},\boldsymbol{\alpha}^{*})=0$. 
In addition, $\hat{H}_{\text{E}}^{Q}(\boldsymbol{\alpha},\boldsymbol{\alpha}^{*})$ contains information 
about the local curvature of the QEL at the critical points and provides the first quantum correction 
to the mean-field approach~\cite{Brandes}. A similar analysis was described in the context of the energy landscape for an ensemble of three-level systems in $\Lambda$-configuration, which are collectively coupled to two bosonic modes~\cite{Hayn2011,Hayn2012}.

%
%%%%%%%%%                                                                  
\subsection{Explicit calculation of the DOQS}
\label{sec:IId}
%%%%%%%%%
%
After the bosonization procedure, one can use the machinery of coherent states~\cite{Glauber} to calculate the traces $\mathcal{T}_{n}$ and the DOQS analytically.

The operator Eq.~\eqref{DisplacementOperator} also allows one to generate
bosonic coherent states 
$\ket{\sqrt{\mathcal{N}}\boldsymbol{\alpha}}=\hat{D}(\sqrt{\mathcal{N}}\boldsymbol{\alpha})\ket{\boldsymbol{0}}$, where $\ket{\boldsymbol{0}}=\ket{0,0,\dots,0}$ is the vacuum state of the bosonic operators~\cite{Glauber}. Consequently, the traces $\mathcal{T}_{n}$ in Eq.~\eqref{DOQSDef} can be easily calculated in the basis $\{\ket{\sqrt{\mathcal{N}}\boldsymbol{\alpha}}\}$ of bosonic coherent states 

\begin{align}
      \label{SCSTrace}
            \mathcal{T}_{n}&=\text{tr}\hat{\mathcal{F}}^{n}=\left(\frac{\mathcal{N}}{\pi}\right)^f\int d^{2f}\boldsymbol{\alpha} \bra{\sqrt{\mathcal{N}}\boldsymbol{\alpha}}\hat{\mathcal{F}}^{n}\ket{\sqrt{\mathcal{N}}\boldsymbol{\alpha}} \nonumber\\
            &=\left(\frac{\mathcal{N}}{\pi}\right)^f\int d^{2f}\boldsymbol{\alpha} \bra{\boldsymbol{0}}\hat{D}^{\dagger}(\sqrt{\mathcal{N}}\boldsymbol{\alpha})e^{-\mathrm{i}n\hat{H}_{\text{E}}T}\hat{D}(\sqrt{\mathcal{N}}\alpha)\ket{\boldsymbol{0}} \nonumber\\
            &=\left(\frac{\mathcal{N}}{\pi}\right)^f\int d^{2f}\boldsymbol{\alpha} \bra{\boldsymbol{0}}e^{-\mathrm{i}n\hat{H}^{(\boldsymbol{\alpha})}_{\text{E}}T}\ket{\boldsymbol{0}} \nonumber\\
            &\approx\left(\frac{\mathcal{N}}{\pi}\right)^f\int d^{2f}\boldsymbol{\alpha} e^{-\mathrm{i} n\mathcal{N} E_{G}(\boldsymbol{\alpha},\boldsymbol{\alpha}^{*})T}F_{n}(\boldsymbol{\alpha} ,\boldsymbol{\alpha}^{*})
       ,
\end{align}
%%%
where $\hat{H}^{(\alpha)}_{\text{E}}$ is the Hamiltonian of Eq.~\eqref{MeanFieldEffHam}.
Furthermore, in Eq.~\eqref{SCSTrace} we have defined a kernel
%%%
\begin{equation}
      \label{QuantumKernel}
            F_{n}(\boldsymbol{\alpha},\boldsymbol{\alpha}^{*})=\bra{\boldsymbol{0}}e^{-\mathrm{i}n\left[\sqrt{\mathcal{N}}\ \hat{H}_{E}^{L}(\boldsymbol{\alpha},\boldsymbol{\alpha}^{*})+\hat{H}_{E}^{Q}(\boldsymbol{\alpha},\boldsymbol{\alpha}^{*})\right]T}\ket{\boldsymbol{0}} ,
\end{equation}
%%%
containing quantum contributions of order $1/\mathcal{N}$.

As in our previous work~\cite{KickedTopESQPT}, we calculate the trace of Eq.
\eqref{SCSTrace} in the thermodynamic limit $\mathcal{N}\gg 1$ by means of the stationary-phase approximation \cite{HaakeBuch}.
Thereby, the trace reads
%%%
\begin{equation}
      \label{SaddlePoinTrace}
            \mathcal{T}_{n}=\frac{1}{n^f}\sum_{\boldsymbol{\alpha}_c\in \mathcal{C}} \frac{\left(\frac{2}{T}\right)^f F_{n}(\boldsymbol{\alpha}_{c},\boldsymbol{\alpha}^{*}_{c})e^{\mathrm{i}\beta_{c}\pi/4}e^{-\mathrm{i} n \mathcal{N}E_{G}(\boldsymbol{\alpha}_{c},\boldsymbol{\alpha}_{c}^{\ast})T}}{\sqrt{|\det\left[\boldsymbol{M}_{G}(\boldsymbol{\alpha} ,\boldsymbol{\alpha}^{*})\right]|_{\boldsymbol{\alpha}=\boldsymbol{\alpha}_{c}}}}
      ,
\end{equation}
%%%
where
%%%
\begin{equation}
       \label{HessianMatrix}
       \boldsymbol{M}_{G}(\boldsymbol{\alpha},\boldsymbol{\alpha}^{*}) =\left(%
\begin{array}{ccc}
\frac{\partial^2 E_{G}(\boldsymbol{\alpha},\boldsymbol{\alpha}^{\ast})}{\partial \alpha_{1}^2}  &  \hdots &\frac{\partial^2 E_{G}(\boldsymbol{\alpha},\boldsymbol{\alpha}^{\ast})}{\partial \alpha_{1} \partial \alpha^{\ast}_{f}}\\
\vdots & \ddots & \vdots\\
\frac{\partial^2 E_{G}(\boldsymbol{\alpha},\boldsymbol{\alpha}^{\ast})}{\partial \alpha^{\ast}_{f} \partial \alpha_{1}} & \hdots &\frac{\partial^2 E_{G}(\boldsymbol{\alpha},\boldsymbol{\alpha}^{\ast})}{\partial (\alpha_{f}^{\ast})^2}
\end{array}
\right)
\end{equation}
%%%
is the Hessian matrix of $E_{G}(\boldsymbol{\alpha},\boldsymbol{\alpha}^{*})$. The sum in Eq.~\eqref{SaddlePoinTrace} is over  $\boldsymbol{\alpha}_{c}\in\mathcal{C}$, where $\mathcal{C}$ is the set of critical points  satisfying the conditions $\frac{\partial E_{G}}{\partial \alpha_{l}}|_{\boldsymbol{\alpha}=\boldsymbol{\alpha}_{c}}=\frac{\partial E_{G}}{\partial \alpha_{l}^{\ast}}|_{\boldsymbol{\alpha}^{\ast}=\boldsymbol{\alpha}^{\ast}_{c}}=0$. 
The index $\beta_{c}$ is the difference in the number of positive and negative eigenvalues of the Hessian matrix $\boldsymbol{M}_{G}(\boldsymbol{\alpha},\boldsymbol{\alpha}^{*})$ for a given critical point. 

The stationary-phase approximation also simplifies the kernel in Eq.~\eqref{QuantumKernel}
%%%
\begin{equation}
      \label{StatPhaseQuantumKernel}
            F_{n}(\boldsymbol{\alpha}_c ,\boldsymbol{\alpha}_c^{*})=\bra{\boldsymbol{0}}e^{-\mathrm{i}n\left[\hat{H}_{E}^{Q}(\boldsymbol{\alpha}_c,\boldsymbol{\alpha}_c^{*})\right]T}\ket{\boldsymbol{0}}
      ,
\end{equation}
%%%
because the linear bosonic terms in the argument of the exponential function vanish at the critical points $\boldsymbol{\alpha}_{c}\in\mathcal{C}$. For completeness we have included a calculation of the Kernel $F_{n}(\boldsymbol{\alpha}_c ,\boldsymbol{\alpha}_c^{*})$ for $f=1$ in appendix~\ref{AppendixA}.
In the limit $\mathcal{N}\gg 1$, we can safely neglect the contribution of the
kernel of
Eq.~\eqref{StatPhaseQuantumKernel}, which has order $1/\mathcal{N}$. Therefore, we consider $F_{n}(\boldsymbol{\alpha}_c ,\boldsymbol{\alpha}_c^{*})\approx 1$ for all the critical points
$\boldsymbol{\alpha}_{c}\in\mathcal{C}$. 

After neglecting the quantum kernel of Eq.~\eqref{StatPhaseQuantumKernel}, we are able to get a semiclassical approximation for the DOQS of Eq.~\eqref{DOQSDef}
%%%
\begin{equation}
      \label{QEDensity}
            \rho_{\text{cl}}(\varepsilon)=\frac{1}{2\pi}+{\rm Re}\left \{\sum_{c \in \mathcal{C}} A_{c} \ e^{\mathrm{i}\beta_{c}\pi/4} {\rm Li}_{f}\left[e^{\mathrm{i}(\varepsilon-E_{c})T}\right] \right\}
      \ ,
\end{equation}
%%%
where $E_{c}=\mathcal{N}E_{G}(\boldsymbol{\alpha}_{c},\boldsymbol{\alpha}_{c}^{*})$ and $\text{Li}_{f}(z)=\sum^{\infty}_{n=1}\frac{z^{n}}{n^{f}}$ is the 
polylogarithm~\cite{Abramowitz}. Correspondingly, the amplitudes $A_{c}$ for each critical point 
are given by
%%%
\begin{equation}
      \label{Amplitude}
            A_{c}= \frac{(2/T)^{f}}{\pi M \sqrt{|\det\left[\boldsymbol{M}_{G}(\boldsymbol{\alpha},\boldsymbol{\alpha}^{*})\right]|_{\alpha=\alpha_{c}}}}
      \ .
\end{equation}
%%%
Interestingly, in the general case of $f>1$, the $(f-1)$th derivative of the DOQS 
%%%
\begin{equation}
      \label{DERDOQS}
            \frac{\partial^{f-1} \rho_{\text{cl}}(\varepsilon)}{\partial \varepsilon^{f-1}}=
            {\rm Re}\left \{\sum_{c \in \mathcal{C}} A_{c} \ e^{\mathrm{i}\frac{(f-1)\pi}{2}}\ e^{\mathrm{i}\frac{\beta_{c}\pi}{4}} {\rm Li}_{1}\left[e^{\mathrm{i}(\varepsilon-E_{c})T}\right] \right\}
\end{equation}
%%%
exhibits a logarithmic divergence if for a given $k \in \mathbb{Z}$, the condition $2(f-1)+\beta_c=8k$ is fulfilled. Let us assume that such a condition is satisfied for a particular critical point $\boldsymbol{\alpha}_{c}\in\mathcal{C}$ with quasienergy $E_{c}$. In this case, the $(f-1)$th derivative of the DOQS  scales as
%%%
\begin{equation}
      \label{DERDOQSLogarithmicDiv}
            \frac{\partial^{f-1} \rho_{\text{cl}}(\varepsilon)}{\partial \varepsilon^{f-1}}\approx -A_{c}\log|\varepsilon-\varepsilon_{c}|
            \ ,
\end{equation}
%%%
where $\varepsilon_{c}=E_{c}\ \text{mod}\ \Omega$ is the genuine critical quasienergy.
This is reminiscent of similar results for non-driven systems that show an ESQPT~\cite{Stransky2014}.
We have included a more precise discussion of the derivation in appendix~\ref{AppendixB}.
%
%%%%%%%%%    
\section{Applications for $f=1$}
\label{sec:III}
%%%%%%%%%
%%%%
%%%%
\begin{figure}
\includegraphics[width=0.49\textwidth,clip=true]{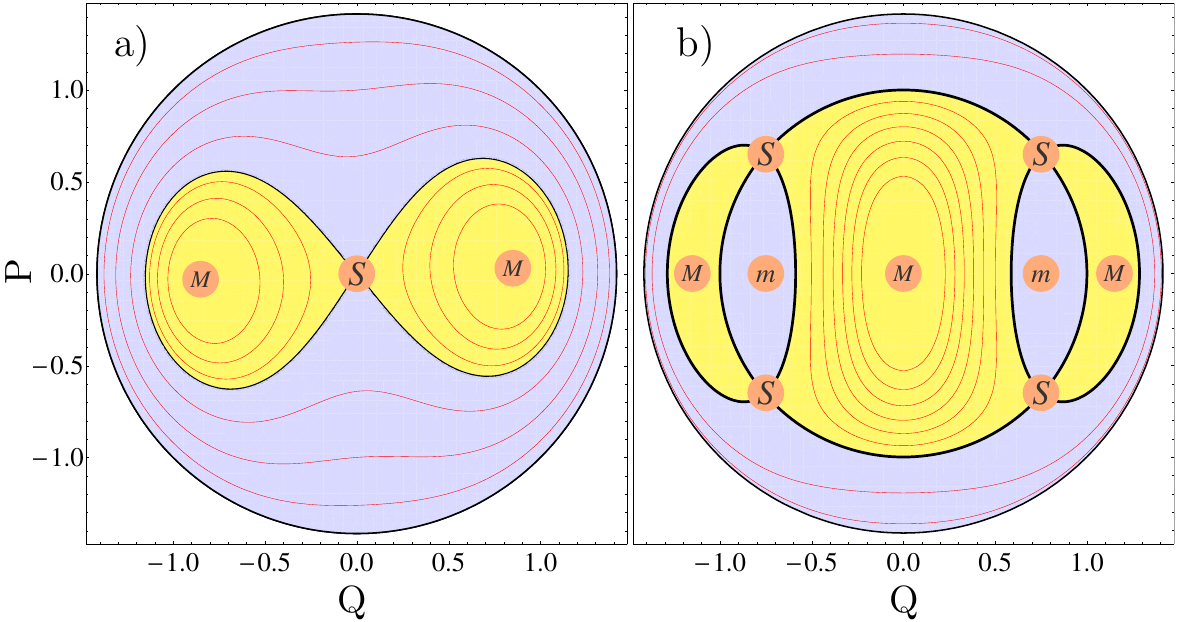}
    \caption{
    (Color online) 
    QEL for (a) the kicked top with $K=0.3$ and (b) the ac-driven model with $GT=20$. The yellow (light gray) and blue (dark gray) regions have energy $E>h$ and $E<h$ for the kicked top, and $E>0$ and $E<0$,  for the ac-driven model, respectively. Solid lines show levels of constant energy. Features are denoted by S (saddle point), M (maxima), and m (minima).
    Other parameters are $\Omega T=2\pi$ and $hT=0.1$.}
    \label{Fig1}
\end{figure}
%%%%
%%%%
To apply the general formalism, in this section we describe the explicit form of the effective Hamiltonians for a fully-connected network of two-level systems 
 with two different kinds of external driving, i.e., delta-kick-type and monochromatic. 
In addition we show the corresponding quasienergy landscapes, calculate the DOQS, and discuss the emergence of critical quasienergy states
and their effects on the magnetization. This work extends and generalizes previous results, published in Ref.~\cite{KickedTopESQPT}.

We assume a fully-connected network of two-level systems with time-dependent interactions given by~\cite{Hangui, Engelhardt2013}
%%%
\begin{eqnarray}
      \label{SpinDriven}
            \hat{H}(t) &=& \frac{h}{2}\sum_{i=1}^{N}\sigma^{(i)}_{x}+
            \frac{g(t)}{4N}\sum_{i,j=1}^{N}\sigma^{(i)}_{z}\sigma^{(j)}_{z} 
\nonumber \\
            &=& hJ_x+\frac{g(t)}{2j} {J_z}^2
      ,
\end{eqnarray}
%%%
where we have defined the collective angular momentum operators $J_{a}=\frac{1}{2}\sum_{i=1}^{N}\sigma^{(i)}_{a}$ with $a \in \{x,y,z,\pm\}$. 
Throughout the paper, we restrict ourselves to the subspace of 
maximally-symmetric states $\ket{j,m}$ with maximal total angular momentum 
$j=N/2$, also known as Dicke states~\cite{Dicke}. 

We consider two different kinds of time-dependent inter-particle interactions, namely delta-kick-type $g(t)=K\sum_{l=-\infty}^{\infty}\delta(t-lT)$ and monochromatic $g(t)=G\cos\Omega 
t$. 
The delta-kick-type driven model is also known as the kicked top~\cite{HaakeBuch,KickedTopESQPT}.

Independent of the chosen driving, in the particular case of Eq.~\eqref{SpinDriven}, the 
Hamiltonian is written in terms of the generators $L_a=J_a$ of the Lie algebra $\mathfrak{g}=su(2)$~\cite{Klein-Marshalek}. Thus, we are strictly limited to the $f=1$ case of the more general theory. 

%
%%%%%%%%%    
\subsection{The effective Hamiltonian and the quasienergy landscape}
\label{sec:IIIa}
%%%%%%%%%
%
Let us first discuss Hamiltonian~\eqref{SpinDriven} with interparticle interaction $g(t)= K \sum_{l=-\infty}^{\infty} \delta(t- l T)$, which corresponds to the kicked top, well studied in the quantum chaos community~\cite{HaakeBuch}. 
In our work, however, to be able to derive the EH one needs to work in the regular regime $hT \sim K \ll 1$ . 
Following the same procedure as in Ref.~\cite{KickedTopESQPT} we obtain the EH
%%%
\begin{align}
      \label{EffectiveHamiltonianKicked}   
            \hat{H}_{\text{E}}
             &=\frac{K}{2j}J^{2}_{z}+\frac{h}{2}\biggl\{\frac{-\mathrm{i}\frac{K}{2j}J_{+}(2 J_{z}+\hat{\mathbbm{1}})}{\exp\left[-\mathrm{i}\frac{K}{2j}(2 J_{z}+\hat{\mathbbm{1}})\right]-\hat{\mathbbm{1}}}+\text{H.c}\biggr\}
      .
\end{align}
%%%
For a more detailed derivation of the EH see appendix~\ref{app:II}. The EH exists as long as we are in the regular regime, as discussed in Ref.~\cite{KickedTopESQPT}.

To obtain the EH for $g(t)=G\cos\Omega t$ in the high frequency limit $h\ll \Omega$ and arbitrary driving amplitude $G$, we consider here a derivation of the EH following 
Refs.~\cite{Hangui,Engelhardt2013}. 
This leads to the EH
%%%
\begin{eqnarray}
            \label{EffectiveHamMonochromatic}
                  \hat{H}_{\text{E}} =\frac{h}{2}J_{+} \mathcal{J}_0\left[\frac{G}{2j\Omega}(2J_z+\hat{\mathbbm{1}})\right] + \text{H.c}\
            ,               
\end{eqnarray}
%%%
where $\mathcal{J}_m(z)$  is the $m$th-order Bessel function~\cite{Abramowitz}.
For completeness, we have included details of the derivation in appendix~\ref{app:II}. 

For both Eqs.~\eqref{EffectiveHamiltonianKicked} and~\eqref{EffectiveHamMonochromatic}, the bosonization procedure can be carried out by means of the Holstein-Primakoff 
representation of the angular momentum operators~\cite{HolsteinPrimakoffOriginal}. In order to provide a geometrical picture, it is convenient 
to define the  coordinates $(X,Y,Z)=(J_{x}/j,J_{y}/j,J_{z}/j)$, which commute in the thermodynamic limit. 
Once we perform the Holstein-Primakoff and 
the shift transformation (for $f=1$) given by Eq.~\eqref{HolsteinPrimakoff} and Eq.~\eqref{DisplacementOperator} respectively, we can write

%%%
\begin{align} 
      \label{ClassicalCoordinates}
            X &=\frac{J_{x}}{j}= 1-\alpha^{*}\alpha    ,  \nonumber \\
            Y &=\frac{J_{y}}{j}=\frac{\alpha^{*}-\alpha}{2\mathrm{i}} \sqrt{2-\alpha^{*}\alpha}   ,   \nonumber   \\
            Z &=\frac{J_{z}}{j}=\frac{\alpha^{*}+\alpha}{2} \sqrt{2-\alpha^{*}\alpha}
     \ ,  
\end{align}
%%%
for $j\gg 1$. To simplify the notation, we have dropped the index of the mean fields defined in Eq.~\eqref{ActionDisplacement}. In addition, due to the conservation of the angular momentum, Eq.~\eqref{ClassicalCoordinates} is the parametrization of the unit sphere in $\mathbb{R}^3$, i.e., the Bloch sphere~\cite{Engelhardt2013}.
%
%%%%%%%%%
\subsection{Discussion of the QEL}
\label{sec:IIIb}
%%%%%%%%%
%

Now we are able to obtain the QEL for the two cases we are interested in. The QEL for the delta-kick-type driving reads
%%%%
\begin{equation}
       \label{QELKicked}
             E_{G}(\alpha,\alpha^{*})=\frac{K}{2}Z^2+\frac{h K Z}{2}\biggl[X\cot\left(\frac{K Z}{2}\right)-Y \biggr]
       ,
\end{equation}
%%%
and for monochromatic driving we obtain
%%%
\begin{equation} 
      \label{QELMonochromatic}
            E_{G}(\alpha,\alpha^{*})=hX\mathcal{J}_0\left(\frac{G}{\Omega}Z\right) 
     .   
\end{equation}
%%%
Figure~\eqref{Fig1} depicts the isocurve values of the energy landscapes of Eqs.~\eqref{QELKicked} and~\eqref{QELMonochromatic}. 
It is worth to mention that an equivalent result can be obtained by using spin coherent
states~\cite{KickedTopESQPT}.

The QEL for the kicked top, unlike for the ac-driving case, exhibits singularities at mean-field level,
along the isocurve values $K Z=2l\pi$ with $l \in \mathbb{Z}$. 
This fact implies that as long as we are far away from the chaotic regime, the QEL is well defined~\cite{KickedTopESQPT}.

Instead of using 
variables $\alpha$ and $\alpha^{*}$, it is more convenient to work with real and imaginary part of $\alpha=Q+\mathrm{i}P$, respectively. 
The benefit of these variables is  that one can depict the Bloch sphere in a restricted domain $Q^2+P^2\le2$ at once, without
splitting the surface in two parts, as
required when using, e.g., stereographic projection. The north pole of the Bloch sphere $(X,Y,Z)=(1,0,0)$ is mapped onto the origin $(Q,P)=(0,0)$, while the south pole of the Bloch sphere
$(X,Y,Z)=(-1,0,0)$ is mapped to the boundary of the domain, i.e., the points $(Q,P)$ such that $Q^2+P^2=2$.

Panel a) in Figure~\ref{Fig1} depicts the QEL of the kicked top model.   There we can find two degenerated maxima $M_{1},M_{2}$, one saddle point $S$ and a minimum $m$ at the boundary of the domain.
We also represent with two different colors the regions divided by the separatrix, which is a curve with constant quasienergy, defined by $E_G(\alpha_{S},\alpha^{\ast}_{S})=h$ , where $E_G(\alpha_{S},\alpha^{\ast}_{S})$ is the quasienergy corresponding to the saddle point  $(Q_{S},P_{S})$. Furthermore,   the separatrix divides the region of the QEL where the 
trajectories are connected, from the region of the QEL where they are not.

Panel b) in Fig.~\ref{Fig1} depicts the QEL  and the isocurve values for the ac-driven  model. In this case, we find three maxima $M_{1},M_{2},M_{3}$, four degenerated 
saddle points $S_{1},S_{2},S_{3},S_{4}$ with energy $E_G(\alpha_{S_{i}},\alpha^{\ast}_{S_{i}})=0$,  and three minima $m_{1},m_{2},m_{3}$, including the boundary of the domain.  We represent the regions divided by the separatrix defined by $E_G(\alpha,\alpha^{\ast})=0$
with different colors.

\subsection{Critical quasienergy states}
\label{sec:IIIc}
%%%%%%%%%
%%%%
%%%%
\begin{figure}
\includegraphics[clip=true,width=0.51\textwidth]{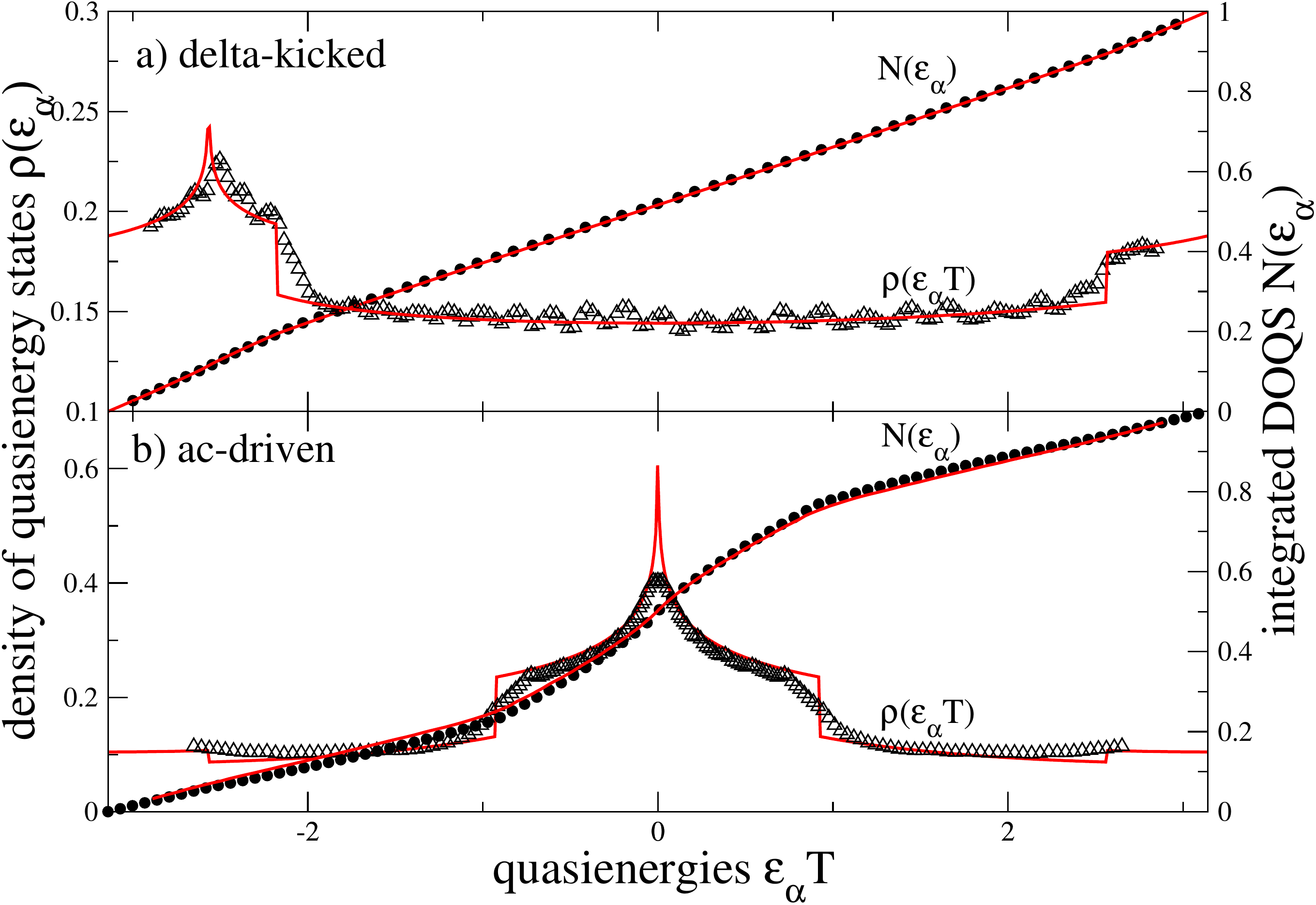}
    \caption{
    (Color online) DOQS $\rho(\varepsilon_\alpha)$ and integrated DOQS $N(\varepsilon_\alpha)$ of (a) the kicked top with $K=0.3$ and  $j=100$ and (b) the ac-driven model with $GT=20$ and  $j=100$
    calculated analytically (red, solid lines) and numerically exact (black symbols). Other parameters are $\Omega T=2\pi$ and $hT=0.1$.}
    \label{Fig2}
\end{figure}
%%%%
%%%%
We have now all the necessary ingredients to calculate the DOQS given by the general formula Eq.~\eqref{QEDensity} with $f=1$ for the QELs of Eqs.~\eqref{QELKicked} and~\eqref{QELMonochromatic}.
Similar to our previous work~\cite{KickedTopESQPT}, in this case $\beta_{M}=2$ at the maxima and  $\beta_{m}=-2$ at the minima, whereas 
$\beta_{S}=0$ for a saddle point. To calculate the DOQS given by Eq.~\eqref{QEDensity}, in the case of delta-kick-type driving one needs to sum over the critical points $\mathcal{C}=\{\alpha_{M_1},\alpha_{M_2},\alpha_{m_1},\alpha_{S}\}$, while for monochromatic driving one has to sum over ten critical points $\mathcal{C}=\{\alpha_{M_{1-3}},\alpha_{m_{1-3}},\alpha_{S_{1-4}}\}$.
%%%

Figure~\ref{Fig2} shows the good agreement between the exact numerical calculation of the DOQS (black triangles and circles) and the analytical result given by
Eq.~\eqref{QEDensity} (red lines) for (a) the kicked top and (b) the ac-driven model. For the sake of completeness, Fig.~\ref{Fig2} also depicts the integrated DOQS 
%%%
\begin{equation}
      \label{IntegratedDOQS}
            N(\varepsilon)=\int^{\varepsilon}_{-\Omega/2}\rho(\epsilon) \ d\epsilon 
      .
\end{equation}
%%%
This quantity inherits the features of the DOQS, which are reflected in a 
discontinuous change of slope at the critical quasienergies.

As a general feature, we find that the saddle 
points $(Q_{S},P_{S})$ of the QEL with quasienergies $E_G(\alpha_{S},\alpha^{\ast}_{S})$ lead to logarithmic-type singularities in the DOQS at critical genuine quasienergies $\varepsilon_{S}=E_{S}\ \text{mod} \  \Omega$, where $E_{S}=j E_G(\alpha_{S},\alpha^{\ast}_{S})$. In the case of delta-kick-type driving, the quasienergy of the saddle point $(Q_{S},P_{S})=(0.0)$ reads $E_G(0,0)T=hT=0.1$, for the parameters of Fig.~\ref{Fig1}~a). This implies that if we choose $j=100$, the singularity must appear at the critical quasienergy
$\varepsilon_{S}T\approx -2.56$, as can be seen in Fig.~\ref{Fig2}~a). In addition, in the case of monochromatic driving, the quasienergy of the saddle points is $E_G(0,0)=0$, as in Fig.~\ref{Fig1}~b), which leads to the singularity located at the quasienergy $\varepsilon_{S}T=0$ in Fig.~\ref{Fig2}~b). 

The singularities previously discussed emerge as a consequence of a clustering of levels in 
the quasienergy spectrum of the system~\cite{KickedTopESQPT}. 
This behavior is characteristic for undriven systems which undergo second-order ESQPTs~\cite{Cejnar06, Caprio08, CejnarStransky08, Leyvraz05}. This leads to the concept of critical quasienergy states (CQS)
for driven systems, which are the natural generalization of ESQPTs to driven  quantum systems.
These CQS are the quantum manifestation of the separatrix  defined by $E_G(\alpha,\alpha^{\ast})=E_G(\alpha_{S},\alpha^{\ast}_{S})$, which is depicted in Fig.~\ref{Fig1}. 

The jumps in the DOQS occur at the genuine quasienergies $\varepsilon_{M}$ and $\varepsilon_{m}$ associated with the maxima and minima, respectively. 
We note that in undriven systems the jumps in the density of states are directly related to first order ESQPTs \cite{CejnarStransky08}, but in the case of external driving, they emerge as a consequence of the periodicity of the quasienergies.

%%%%%%%%%
\subsection{Signatures of critical quasienergy states arising in observables of the system}
\label{sec:IIId}
%%%%%%%%%
It is well known that singular behavior of the density of states in undriven systems is also reflected in observables of the system~\cite{RibeiroESQPT, BrandesESQPT}. 
In a similar fashion, under the effect of external control, CQS should also appear in observables of the system, as they can be expressed in terms of derivatives of the DOQS~\cite{KickedTopESQPT}. This is a direct consequence of the extension of the Hellmann-Feynman 
theorem to Floquet theory~\cite{HanggiGrifoni}. 

While the DOQS is not very well accessible experimentally, the magnetization has already been measured in driven cold-atom experiments~\cite{Zibold2010,Gross2010,Chaudhury1,Chaudhury2}.
We thus focus in the following on the scaled transverse magnetization $\expval{J_x/j}$. It is convenient to define the expectation value using the quasienergy eigenstates,
\begin{equation}
 \expval{J_{x}/j}_{\mu} \equiv \bra{\Phi_{\mu}(0)} \frac{J_x}{j} \ket{\Phi_{\mu}(0)}\,,
\end{equation}
where $\ket{\Phi_{\mu}(0)}$ is the 
Floquet mode with quasienergy $\varepsilon_{\mu}$. 

\begin{figure}[h]
\includegraphics[width=\linewidth,clip=true]{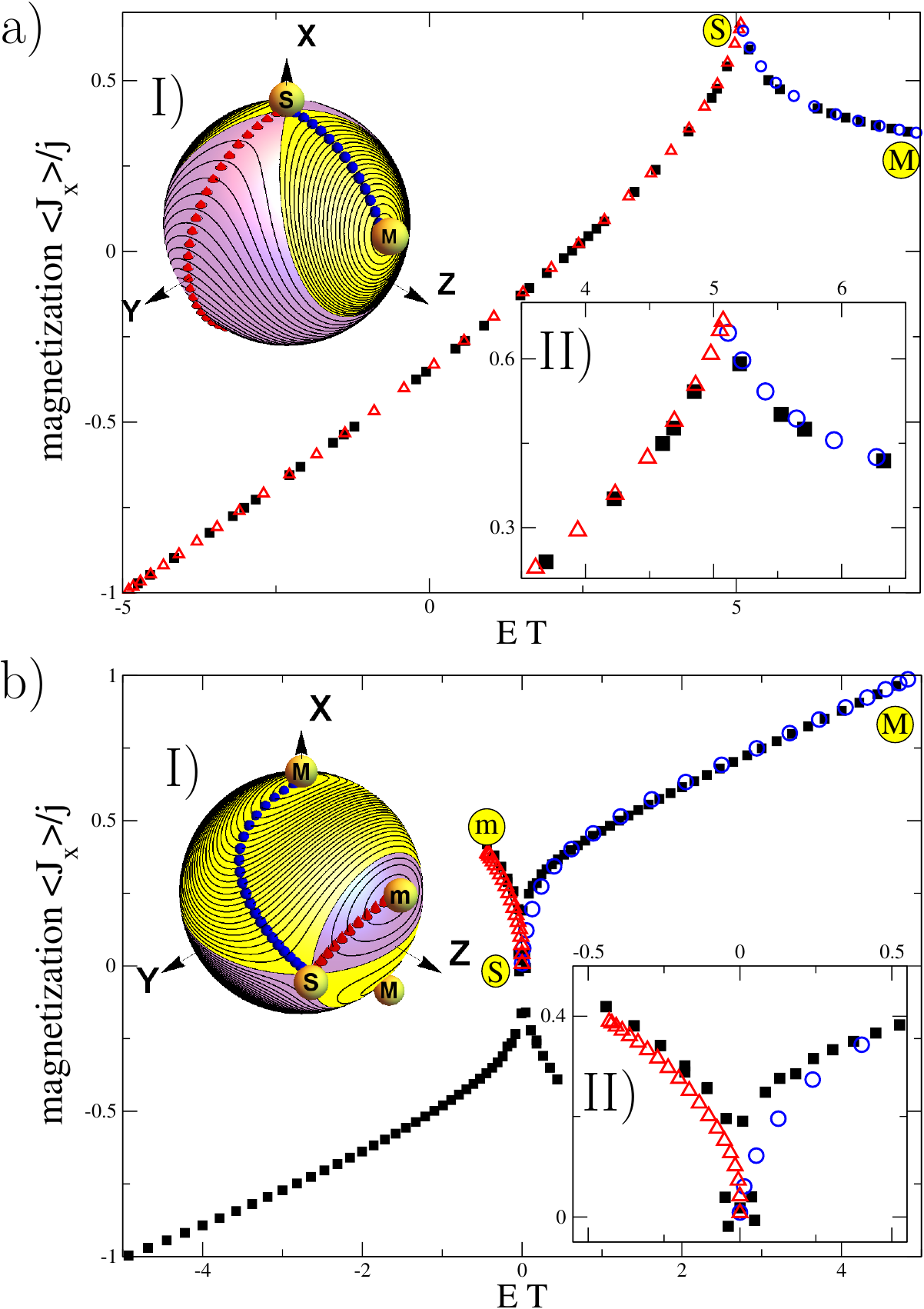}
    \caption{
   (Color online) Magnetization $\langle J_{x}/j\rangle_{\mu}$ for (a) the kicked top with $K=0.3$, and (b) the ac-driven LMG model with $GT=20$.
   Calculations in quasienergy states (filled, black squares) are compared to time-averaged expectation values (blue circles and red triangles) with initial conditions shown in Insets (I), respectively. 
   Insets (II) show details of the cusp arising due to the CQS for the different drivings. Features of the corresponding QEL are denoted by S (saddle point), M (maxima), and m (minima).
   Other parameters are $j=50$, $\Omega T =2\pi$, and $hT=0.1$.
    }
   \label{Fig3}
\end{figure}

However, from an experimental point of view, it is challenging to prepare the system in a given Floquet mode
$\ket{\Phi_{\mu}(0)}$. 
For this reason, similarly to Refs.~\cite{KickedTopESQPT,Engelhardt2014}, we propose here a measurement protocol to observe the cusp behavior in the transverse magnetization.
To initialize the measurement, we propose to prepare the system in a spin coherent state 
$\ket{\Psi(0)}=\ket{\gamma}$ following the definition of Ref.~\cite{NoriSpinSqueez}
%%%
\begin{equation}
      \label{DefSpinCoherentStates}
            \ket{\gamma}=(1+\gamma\gamma^{\ast})^{-j} e^{\gamma (J_{z}-iJ_{y})}\ket{j,j}_{x}
      \ ,
\end{equation}
where $\ket{j,j}_x$ denotes Dicke states in the $J_x$-basis. 
%%%
We choose the spin coherent state to be centered 
at \mbox{$\boldsymbol{R}_0=[X_0,Y_0,Z_0]$} on the Bloch sphere~\cite{KitagawaUeda,NoriSpinSqueez}, in such a way that
%%%
\begin{equation}
      \label{InitialConditionsStereograph}
            \gamma=\frac{Z_0}{1+X_0}+\mathrm{i}\frac{Y_0}{1+X_0}
      \ .
\end{equation}
%%%
The insets (I) in Figure~\ref{Fig3} depict the initial conditions $\boldsymbol{R}_0$ for the measurement protocol both for (a) delta-kick type driving and (b) monochromatic driving.
Blue circles show initial conditions between saddle point (S) and maximum (M), while red triangles denote initial conditions between saddle point and minimum (m).
%
%
%%%%
%%%%
%%%%
%%%%
Given an experimental register of the stroboscopic evolution $\ket{\Psi(lT)}=\mathcal{F}^{l}\ket{\Psi(0)}$ during $L$ periods 
of the driving -- cf. Eq.~\eqref{EvolutionStroboscopic} --, it
is natural to define the time-averaged density operator 
%%%
\begin{equation}
      \label{TimeAveragedDensityOperator}
            \overline{\rho}=\frac{1}{L+1}\sum^{L}_{l=0}\ket{\Psi(lT)}\bra{\Psi(lT)}.
\end{equation}
%%%
Correspondingly, the time-averaged expectation value of an observable $\hat{\mathcal{O}}$ reads
$\overline{\expval{\hat{\mathcal{O}}}}=\rm{tr}(\overline{\rho}\hat{\mathcal{O}})$.
In our measurement protocol, the initial state has a fixed unfolded quasienergy
$\bra{\Psi(0)}\hat{H}_{\text{E}}\ket{\Psi(0)}=E_{\mu}$, which remains constant after the time average
$\overline{\expval{\hat{H}_{\text{E}}}}=E_{\mu}$. 
Now one can plot the coordinates $(E_{\mu},\overline{\expval{J_{x}/j}})$ to compare with the result in eigenstates
$(E_{\mu},\expval{J_{x}/j}_{\mu})$ as it is shown in Fig.~\ref{Fig3}. 

The filled symbols in Fig.~\ref{Fig3} depict the expectation value of the scaled magnetization in Floquet modes for a finite system size $j=50$ for (a) delta-kick-type driving and (b) ac-driving as a function of the unfolded quasienergies $\langle\hat{H}_{\text{E}}\rangle_{\mu}=E_{\mu}$. We note that a cusp behavior of the magnetization appears at the critical quasienergy $E_{S}T=jhT=5$ for the kicked top, and $E_{S}T=0$ for the ac-driven  model. This result follows
directly from the behavior of the DOQS observed in Figs.~\ref{Fig2}~a)~and~b).
A similar behavior of this observable has been found in undriven LMG-type and Dicke-type models~\cite{Caprio08, BrandesESQPT,PPF11A,PPF11E}.

The insets $(I)$ in Fig.~\ref{Fig3} depict the chosen initial conditions on the Bloch spheres. 
For both cases, we select initial conditions  along a path 
joining the minimum with the saddle point, as well as along a path joining the saddle point with a maximum of the QELs depicted in Fig.~\ref{Fig1}.
The points along the paths $\boldsymbol{R}_{0}$ are chosen such that they
exhibit a minimal velocity of the semiclassical system $V(\alpha,\alpha^{\ast})=\sqrt{(\frac{\partial E_{G}}{\partial \alpha})^2+(\frac{\partial E_{G}}{\partial \alpha^{\ast}})^2}$. This leads to a minimal participation ratio of the initial state, which results in
a small deformation of the wave packet during the time evolution~\cite{Scharf1992,Weaire1977,Berry1972}.
In Ref.~\cite{Engelhardt2014} this relation is discussed in more detail for the undriven 
LMG model. 
Additionally, the open symbols in Fig.~\ref{Fig3} b) show the results of the measurement protocol only for the upper branch of the transverse magnetization. 
The lower branch can be obtained by considering the symmetry transformation $J_{x}\rightarrow-J_{x}$ and $E_{\mu}\rightarrow-E_{\mu}$.
The insets $(II)$ show a zoom into the cusp region and underscore the good agreement of the measurement protocol with the result for quasienergy eigenstates.

%%%%%
\section{Conclusions}
\label{sec:IV}
%%%%%

We have studied signatures of CQS in mean-field-type spin models under delta-kick-type and monochromatic driving. By assuming that it is possible to bosonize the EH of the driven system in terms of $f$ bosonic operators, we have derived a general formula for the DOQS. In the particular case of a fully-connected network of two-level systems with time-dependent interactions, most of the features of the DOQS are due to the nature of the critical points which appear
in the QELs, i.e., saddle points are responsible for logarithmic divergences in the DOQS. Also, we have explored how this CQS
can be observed in the scaled magnetization of the system. For this
purpose, we have developed a measurement protocol to test experimentally CQS in driven systems. This measurement protocol relies 
on the time-averaged expectation values of the system initialized in a  coherent state on the Bloch sphere. We have shown that the quantum signature of the separatrix appears in the cusp behavior of the scaled magnetization, similarly to the cusp that can be found in undriven systems.

Future directions of research include the application of the theory to driven Dicke- and $\Lambda$- models ($f=2,4$)~\cite{Bastidas2012,Hayn2011}, the use of CQS for the generation of squeezed states and quantum metrology~\cite{NoriSpinSqueez}, and the extension to driven-dissipative systems~\cite{Vorberg2013}.

%%%%%%%%%%%%%%%%%%%%%%%%%%%%%%%%%%%%%%%%%%%%%%%%%%%%%%%%%%%%%%%%%%%%%%%%%%%%%%%%%%%%%%%%%%%%%%%%%%%%%%%%%%%%
\begin{acknowledgments}
V.M.B acknowledges inspiring discussions with A. Buchleitner, M. Gessner, C. Nietner, A. Sorokin, and P. Strasberg.
The authors gratefully acknowledge financial
support by the DFG via grants BRA 1528/7, BRA 
1528/8, SFB 910 (V.M.B., T.B.), the Spanish Ministerio de 
Ciencia e Innovaci\'on (Grants No. FIS2011-28738-C02-01) and Junta de 
Andaluc\'ia 
(Grants No. FQM160) (P.P.-F.). 
\end{acknowledgments}
%%%%%%%%%%%%%%%%%%%%%%%%%%%%%%%%%%%%%%%%%%%%%%%%%%%%%%%%%%%%%%%%%%%%%%%%%%%%%%%%%%%%%%%%%%%%%%%%%%%%%%%%%%%%%%%%%%%%%%%%%%%
%
\appendix
%
%%%%%%%%
\section{Calculation of the quantum Kernel $F_{n}(\alpha_c ,\alpha_c^{*})$ \label{AppendixA}}
%%%%%%%% 
Let us consider the quadratic part of the Hamiltonian~\eqref{MeanFieldEffHam} for $f=1$, which posses the canonical form of the squeezing Hamiltonian
%%%
\begin{equation}
      \label{CanonicalSqueezing}
            \hat{H}_{E}^{Q}(\alpha_c,\alpha_c^{*})=\omega_{c} a^{\dagger}a+\Gamma_{c} \left[a^2+ (a^{\dagger})^2\right]
      ,
\end{equation}
%%%
where the parameters $\omega_{c}$ and $\Gamma_{c}$ contain information of the local geometry of the critical points $\alpha_{c}\in \mathcal{C}$.
Let us write the Hamiltonian~\eqref{CanonicalSqueezing} in terms of the quadratures $q=(2\omega_{c})^{-1/2}(a^{\dagger}+a)$ and 
$p=\mathrm{i}(\omega_{c}/2)^{1/2}(a^{\dagger}-a)$ of the  bosonic field as in Ref.~\cite{Brandes}, as follows
%%%
\begin{equation}
      \label{QuadraturesSqueezing}
            \hat{H}_{E}^{Q}(\alpha_c,\alpha_c^{*})=\frac{p^{2}}{2}+\frac{\vartheta_{c}^{2}}{2}q^{2}-\frac{\omega_{c}}{2}
      ,
\end{equation}
%%%
where $\vartheta_{c}^{2}=\omega_{c}^{2}-4\Gamma_{c}^2$. 
The sign of $\vartheta_{c}^{2}$ varies depending on the geometry of the critical points, i.e., $\vartheta_{c}^{2}>0$ 
for maxima (M) and minima (m) and $\vartheta^{2}_{c}<0$ for a saddle point (S). 

In position representation we can write the quantum correction Eq.~\eqref{StatPhaseQuantumKernel} in terms of the propagator $G(x,y;t)$ of the one-dimensional harmonic oscillator~\cite{Brandes}
%%%
\begin{equation}
      \label{QuantumKernelGreen}
            F_{n}(\alpha_c ,\alpha_c^{*})=\int_{-\infty}^{\infty}\int_{-\infty}^{\infty}\ dq\ dq' \ \psi^{\ast}_{0}(q)G(q,q';nT)\psi_{0}(q')
      ,
\end{equation}
%%%
where $\psi_{0}(q)=\braket{q}{0}=(h/\pi)^{1/4}e^{-(h/2)q^{2}}$ and
%%%
\begin{equation}
      \label{SqueezedOscilatorGreen}
            G(q,q';t)=\mathcal{N}_{c}
            \exp\left\{ \frac{\mathrm{i}\vartheta_{c}[(q^2+(q')^2)\cos(n\vartheta_{c}T)-2qq']}{2 \sin(n\vartheta_{c}T)}\right\}
      \ ,
\end{equation}
%%%
with $\mathcal{N}_{c}=\left[2\pi\mathrm{i} \sin(n\vartheta_{c}T)/\vartheta_{c}\right]^{-1/2}$.
Now we proceed to write the quantum correction of Eq.~\eqref{StatPhaseQuantumKernel} in a suggestive
way
%%%
\begin{equation}
      \label{GaussianIntegral}
            F_{n}(\alpha_c ,\alpha_c^{*})=\widetilde{\mathcal{N}_{c}}\int_{-\infty}^{\infty}d^{2}\mathbf{r}\exp\left(-\frac{1}{2}\mathbf{r}^{T}\cdotp\mathbf{A}\cdotp\mathbf{r}\right)=\frac{2\pi\widetilde{\mathcal{N}_{c}}}{\sqrt{\det\mathbf{A}}}
      ,
\end{equation}
%%%
where $\widetilde{\mathcal{N}_{c}}=(h/\pi)^{1/2}\left[2\pi\mathrm{i} \sin(n\vartheta_{c}T)/\vartheta_{c}\right]^{-1/2}$ and $\mathbf{r}^{T}=(q,q')$. 
The matrix representing  the quadratic form in the argument of the exponential reads
%%%
\begin{equation}
      \label{TimeDepBogoliubovHamiltonian}
\mathbf{A}=\left(%
\begin{array}{cc}
h-\mathrm{i}\vartheta_{c}\cot(n\vartheta_{c}T) & \frac{\mathrm{i}\vartheta_{c}}{\sin(n\vartheta_{c}T)}\\
 \frac{\mathrm{i}\vartheta_{c}}{\sin(n\vartheta_{c}T)} &  h-\mathrm{i}\vartheta_{c}\cot(n\vartheta_{c}T)
\end{array}
\right)
\ .
\end{equation}
%%%
Finally, we can write 
%%%
\begin{equation}
      \label{GaussianIntegralApp}
            F_{n}(\alpha_c ,\alpha_c^{*})= \sqrt{\frac{-2\mathrm{i}h\vartheta_{c}\sin(n\vartheta_{c}T)}{[h\sin(n\vartheta_{c}T)-\mathrm{i}\vartheta_{c}\cos(n\vartheta_{c}T)]^{2}+\vartheta_{c}^{2}}}
      .
\end{equation}
%%%
%%%%%%%%%%%%%%%%%%%%%%%%%%%%%%%%%%%%%%%%%%%%%%%%%%%%%%%%%%%%%%%%%%%%%%%%%%%%%%%%%%%%%%%%%%%%%%%%%%%%%%%%%%

%%%%%%%%
\section{Detailed study of the DOQS \label{AppendixB}}
%%%%%%%%
In this appendix we discuss in more detail the derivation of Eq.~\eqref{DERDOQS} in the main text.
For an arbitrary integer number $f$, the DOQS given in Eq.~\eqref{QEDensity} has interesting properties. 
Let us begin by considering the identity 
%%%
\begin{equation}
      \label{IdentityPoly}
            \frac{\partial^{r}}{\partial \theta^{r}}\text{Li}_f(e^{\mathrm{i}\theta})=\mathrm{i}^{r}\text{Li}_{f-r}(e^{\mathrm{i}\theta})
\end{equation}
%%%
satisfied by the polylogarithm $\text{Li}_f(e^{\mathrm{i}\theta})$~\cite{Abramowitz}. 
As a consequence of this, if one calculates the $(f-1)$-th derivative of the DOQS given in Eq.~\eqref{QEDensity} with respect to the quasienergy $\varepsilon$, one obtains Eq.~\eqref{DERDOQS}.

Motivated by a previous work~\cite{KickedTopESQPT}, we can use the expansion of polylogarithm~\cite{Abramowitz}
%%%
\begin{equation}
      \label{PolyLog}
            {\rm Li}_{1}\left(e^{\mathrm{i}\theta}\right)
            =-\log\left[2\sin\left(\frac{\theta}{2}\right)\right]+\mathrm{i}\left(\frac{\pi-\theta}{2}\right)
     ,
\end{equation}
%%%
where $0\leq\theta<2\pi$.
From Eq.~\eqref{PolyLog} follows that if $2(f-1)+\beta_c=8k$ for $k \in \mathbb{Z}$, the DOQS exhibits a logarithmic divergence as in Eq.~\eqref{DERDOQSLogarithmicDiv}. In the particular case of $f=1$, one obtains $k=0$ when one evaluates the index $\beta_{S}=0$ for a saddle point $S$. In the case of maxima $M$ and minima $m$ does not exists an integer $k$ such that $\beta_{M,m}=8k$ because $\beta_{M}=2$ and $\beta_{m}=-2$~\cite{KickedTopESQPT}. Therefore, in this case the DOQS exhibits jumps at the  quasienergies $\varepsilon_{M}$ and $\varepsilon_{m}$.

%%%%%%%%%%%%%%%%%%%%%%%%%%%%%%%%%%%%%%%%%%%%%%%%%%%%%%%%%%%%%%%%%%%%%%%%%%%%%%%%%%%%%%%%%%%%%%%%%%%%%%%%%%%%
\section{Derivation of the effective Hamiltonians\label{app:II}}
%%%%%%%%%
Our first step is to show how to derive the EH for a delta-kick-type modulation of the inter-particle interaction 
$g(t)=K\sum_{l=-\infty}^{\infty}\delta(t-lT)$ in Eq.~\eqref{SpinDriven}.
Working in the regular regime of the kicked-top, within one period,
the propagator factorizes into two parts
%%%
\begin{equation}
      \label{FloqOpKick}
	    \hat{\mathcal{F}}=e^{-\mathrm{i} hT J_{x}} e^{-\mathrm{i} (K/2j) J^{2}_{z} }
      \ .
\end{equation}
%%%
Following the same procedure as in Ref.~\cite{KickedTopESQPT}, we use the Baker-Campbell-Hausdorff (BCH) formula in the regime
$hT \sim K\ll 1$ to construct  $\hat{H}_{\text{E}}$. With this aim, we use that the Floquet operator Eq.~\eqref{FloqOpKick} can be written in the form $\hat{\mathcal{F}}=e^{-\mathrm{i}h\hat{B}}e^{-\mathrm{i}\hat{A}}=e^{-\mathrm{i}\hat{H}_{\text{E}}}$ with $\hat{A}=\mathrm{i}\frac{K}{2j}J^{2}_{z}$ and $\hat{B}=\mathrm{i}T J_{x}$. The BCH formula allows one to obtain the EH
%%%
\begin{align}
      \label{EffectiveHamiltonianKicked}   
            \hat{H}_{\text{E}}
             &=  -\mathrm{i}\hat{A}+\mathrm{i} h \frac{\text{ad}_{\hat{A}}}{\exp[-\text{ad}_{\hat{A}}]-\hat{\mathbbm{1}}}\hat{B}\nonumber
      ,
\end{align}
%%%
where $\text{ad}_{\hat{X}}\hat{Y}=[\hat{X},\hat{Y}]$ denotes the adjoint 
representation of the angular momentum algebra~\cite{Scharf1}. This finally leads so expression~\eqref{EffectiveHamiltonianKicked}. 

To derive the EH for $g(t)=G\cos\Omega t$ we require to construct the evolution operator in one
period of the driving for the Hamiltonian~\eqref{SpinDriven}.
We consider here a derivation of the EH following 
Refs.~\cite{Hangui,Engelhardt2013}. To accomplish this task, we work in the 
interaction picture, in which the Floquet operator reads
$\hat{\mathcal{F}}=\hat{U}_{0}(T)\hat{U}_{I}(T)$, where $\hat{U}_{0}(t)=\exp\left(-\mathrm{i}\frac{G\sin\Omega t}{2j \Omega} {J_z}^2\right)$ with $\hat{U}_{0}(T)=\hat{\mathbbm{1}}$, and 
%%%
\begin{equation}
      \label{EvolInteraction}
            \hat{U}_{I}(t)=\hat{\mathcal{T}}\exp\left(-\mathrm{i}h\int^{t}_{0}\hat{U}^{\dagger}_{0}(\tau)\ J_{x}\ \hat{U}_{0}(\tau)\ d\tau\right)
\end{equation}
%%%
is the evolution operator in the interaction picture. 
In the high-frequency limit $h\ll\Omega$, one can expand the Floquet operator 
as follows
%%%
\begin{align}
            \label{FloquetMonochromatic}
                 \hat{\mathcal{F}}&=\hat{U}_{I}(T)\approx\hat{\mathbbm{1}}-\mathrm{i}h\int^{T}_{0}\hat{U}^{\dagger}_{0}(\tau)\ J_{x}\ \hat{U}_{0}(\tau)\ d\tau \nonumber\\
                 &= \hat{\mathbbm{1}}-\mathrm{i}\frac{h}{2}
                 \left[J_{+}\int^{T}_{0}e^{-\mathrm{i}\frac{G\sin\Omega \tau}{2j\Omega} (2J_{z}+\hat{\mathbbm{1}})} d\tau + \text{H.c}\right]
            .               
\end{align}
%%%
By using the expansion $e^{\mathrm{i}z\sin \Omega t}=\sum^{\infty}_{m=-\infty}\mathcal{J}_m(z)e^{\mathrm{i}m\Omega t}$,
where $\mathcal{J}_m(z)$  is the $m$th-order Bessel function~\cite{Abramowitz}, one can express approximately the last line in terms of an exponential
$\hat{\mathcal{F}}\approx e^{-\mathrm{i}\hat{H}_{\text{E}}T}$, which leads to the EH in Eq.~\eqref{EffectiveHamMonochromatic}.
%%%
%%%%%

%%%%%%%%%%%%%%%%%%%%%%%%%%%%%%%%%%%%%%%%%%%%%%%%%%%%%%%%%%%%%%%%%%%%%%%%%%%%%%%%%%%%%%%%%%%%%%%%%%%%%%%%%%%%%%%%%%%%%%%%%%%%%%%%%%%%%%%

\end{document}